\documentclass[english,twocolumn,showpacs,preprintnumbers,amsmath,amssymb,showkeys]{revtex4-2}
\usepackage{graphicx}% Include figure files
\usepackage{dcolumn}% Align table columns on decimal point
\usepackage{bm}% bold math
\usepackage{bigints}
\usepackage{babel,amssymb,amsmath,graphicx,hyperref}

\begin{document}
\title{Observing an accretion disk inside a wormhole shadow}

\author{I.D. Novikov}
\affiliation{Astro-Space Center of P.N. Lebedev Physical Institute, Profsoyusnaya 84/32, Moscow, Russia 117997,}
\affiliation{The Niels Bohr International Academy, The Niels Bohr Institute, Blegdamsvej 17, DK-2100, Copenhagen, Denmark}
\affiliation{National Research Center Kurchatov Institute, 1, Akademika Kurchatova pl., Moscow,  Russia 123182}
\author{S.V. Repin}
\affiliation{Astro-Space Center of P.N. Lebedev Physical Institute, Profsoyusnaya 84/32, Moscow, Russia 117997}
\author{D.A. Paksivatova}
\affiliation{Physics and Mathematics Lyceum No. 2007, 9, Gorchakova str., Moscow, 117042, Russia}

\begin{abstract}
The paper considers the problem of the possibility of observing an accretion disk through the throat of the 
Ellis-Bronnikov-Morris-Thorne wormhole. It is shown that this image has a complex structure and is fundamentally 
different from the image of an accretion disk around a black hole. Images of the accretion disk are presented at 
various angles between the plane of the disk and the observer's line of sight.
\end{abstract}


\maketitle

\section{Introduction}

       One of the most surprising predictions of the general theory of relativity is the prediction of the possibility of the existence 
of wormholes \cite{Einstein_1935, 1970eivi.book.....W, Ellis_1973, Bronnikov_1973, Morris_1988a, Morris_1988b}. In the simplest 
models, wormholes (WH) are  tunnels connecting two asymptotically flat regions of space-time. The tunnel itself lies outside 
our space-time. Such WH are static and do not change over time. Both entrances and exits of the tunnel are similar to black 
holes in their external geometric properties \cite{2021ARep...65....1N}. The most significant difference between them and black 
holes is that the light signals can enter and exit them, passing through the tunnel. Thus, from one asymptotically flat region it is, 
in principle, possible to observe another. Such objects may exist in our Universe. Seaching them using the latest methods of 
space astrophysics is a major breakthrough task~\cite{Novikov_2021}. For these purposes, it is necessary to know the distinctive 
observational features of wormholes and how they differ from the observational properties of black holes.

      Like black holes, the entrances to wormholes can be surrounded by the accretion disks. In the case of a wormhole, an 
observer located in another asymptotically flat region of space can see the disk through the tunnel due to the rays passing 
through the mouth of a wormhole.

      The objective of this work is to identify the fundamental features of the visibility of the disk observed through the throat 
of a wormhole. Related issues and approaches to solving these problems were considered in the papers 
\cite{2017PhRvD..95b4030T, 2017PhRvD..95f4035T, 2018PhyU...61..280N, Bugaev_2021, 2023PhRvD.108l4059B, Bugaev_2022b, 
Repin_2022, Bugaev_2022c, 2023EPJC...83..284T, 2023EPJC...83..284T, 2024EPJC...84.1325T, 2024EPJC...84..480S, 
2025NuPhB101416876T, 2022PhRvD.105f4013T}. Our approach differs from these works by considering precisely 
the fundamental issues.

\section{Wormhole metric and the geometric parameters of the model}

In the paper we considers the static Ellis-Bronnikov-Morris-Thorne wormhole, the metric of which can be written as
\begin{equation}
     ds^2 = dt^2 - \cfrac{r^2}{r^2 - q^2}\,\, dr^2 - r^2
            \left(
               d\vartheta^2 + \sin^2\vartheta \, d\varphi^2
            \right)\,,
          \label{MT_metric2}
\end{equation}
where the constant $q$ is the radius of the wormhole's throat. In the equation (\ref{MT_metric2}) the speed of light is taken 
to be unity, $c=1$, and the radial coordinate~$r$ is chosen so that the circumference is~$2\pi r$. The coordinate $r$ is 
convenient to use for constructing the trajectories of light rays, because the trajectories of photons then look more natural 
and clear.
        
\begin{figure}[!htb]
  \centerline{
  \includegraphics[width=0.95\columnwidth]{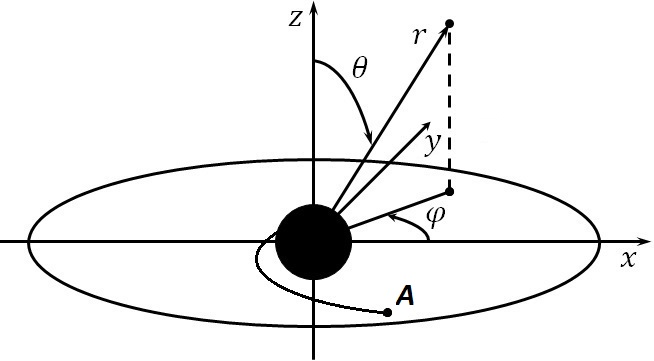}
             }
  \caption{The location of the accretion disk and wormhole. 
                  The geometric parameters of the model and the coordinates used.}
  \label{Disk_and_geometry}
\end{figure}
        
\begin{figure}[!htb]
  \centerline{
  \includegraphics[width=0.95\columnwidth]{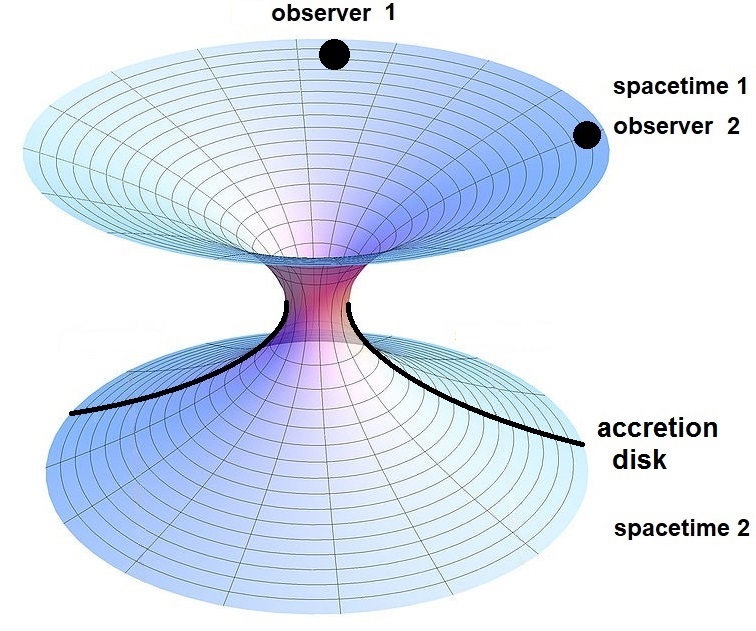}
             }
  \caption{Two-dimensional cross-section of a wormhole. Observer and accretion disk locations. 
                  The observer is in space-1, and the accretion disk is in space-2 and reaches the throat of the wormhole.}
  \label{Observer_and_disk}
\end{figure}

       Fig.~\ref{Disk_and_geometry} and \ref{Observer_and_disk} schematically show the geometric para\-meters of the model. 
Fig.~\ref{Disk_and_geometry} shows the location of the accretion disk near the wormhole. Here the black circle denotes the 
wormhole, and the ellipse denotes the outer boundary of the accretion disk surrounding it. In this paper, we will assume that 
this disk extends from the mouth of the wormhole to infinity. We consider this disk to be thin and homogeneous. This model 
allows us to choose an asymptotically Cartesian coordinate system so that the disk lies in the $xy$-\glqq plane\grqq. 
The direction of the $x$ axis can be chosen arbitrarily. The $z$ axis is directed perpendicular to the disk plane. In the 
Boyer--Lindquist coordinate system, the radial coordinate is denoted by $r$, the azimuthal angle $\varphi$ is traditionally 
measured from the positive direction of the $x$ axis (polar axis), and the polar angle~$\theta$~is measured from the 
perpendicular to the $xy$-\glqq plane\grqq{} of the accretion disk. The value of the $r$ coordinate cannot be less than 
the radius of the wormhole throat, which obviously follows from the formula~(\ref{MT_metric2}). Fig.~\ref{Disk_and_geometry} 
also shows the trajectory of one of the photons that came out of the wormhole's mouth and, after flying some distance, 
reached the accretion disk at point~$A$. The trajectory of such a photon will, of course, be curved.

      Figure~\ref{Observer_and_disk} shows a two-dimensional section of the wormhole, the location of the same disk as shown 
in Figure~\ref{Disk_and_geometry}, and an observer. We will assume that the observer is in space-1 and records the radiation 
of the accretion disk, which passes through the throat of the wormhole from space-2. The position of the observer relative 
to the accretion disk, of course, can change. Figure~\ref{Observer_and_disk} also shows the location of the accretion disk 
in space-2 and two observers in space-1. The coordinate system introduced in space-2, i.e. in the space where the accretion 
disk is located, can be continued into the observer's space. Then, by analogy, we can consider the polar angle $\theta$ 
between the line of sight of the observer in space-1 and the perpendicular to the plane of the accretion disk in space-2. 
Using this analogy, we can say that in Fig.~\ref{Observer_and_disk} the ray going from observer-1 towards the wormhole 
will be \glqq perpendicular\grqq{} to the plane of the accretion disk, and observer-2 is located approximately \glqq in the 
plane of the disk\grqq. This analogy is convenient for understanding the structure of the image.

       In the centrally symmetric Ellis-Bronnikov-Morris-Thorne metric, the principle of ray reversibility is satisfied, so in 
numerical simulation it is more convenient to consider the movement of light rays from the observer to the object, rather 
than from the object to the observer. Both of these trajectories completely coincide if we change the sign of time   and 
thereby reverse the course of the ray.

      In the Ellis--Bronnikov--Morris--Thorne metric, the last stable orbit lies at the wormhole's throat, i.e. at $r=q$. This 
means that the inner boundary of the accretion disk in space-2 must reach the wormhole throat. This situation is 
fundamentally different from the Schwarzschild black hole, where the last stable orbit is at a distance of $r = 3r_g$ 
($r_g = 2Gm/c^2$). And this, in turn, must lead to observational effects that are fundamentally different for these objects.

\section{Equations of a photon motion the wormhole field}

        To construct an image, it is necessary to calculate a large number of photon trajectories (zero geodesics) in the 
Ellis--Bronnikov--Morris--Thorne metric. The equations of photon motion can be reduced to a system of six ordinary 
differential equations \cite{Zakharov_1994, Zakh_Rep_1999, Repin_2022}:
\begin{eqnarray}
      \cfrac{dt}{d\sigma} & = & \cfrac{1}{R^2}\,\,, \label{Eq_motion2_1}  \\
      \cfrac{dR}{d\sigma} & = & R_1\,, \label{Eq_motion2_2}  \\
      \cfrac{dR_1}{d\sigma} & = & 2 \left(\eta - \xi^2\right) R^3 - \left(1 + \eta + \xi^2\right) R\,, 
                                \label{Eq_motion2_3} \\
     \cfrac{d\theta}{d\sigma} & = & \theta_1\,, \label{Eq_motion2_4} \\ 
     \cfrac{d\theta_1}{d\sigma} & = & \cfrac{\xi^2\cos\theta}{\sin^3\theta}\,\,,\label{Eq_motion2_5} \\
      \cfrac{d\varphi}{d\sigma} & = & \cfrac{\xi}{\sin^2\theta}\,\,.  \label{Eq_motion2_6}
\end{eqnarray}
where $t, r = 1/R, \theta, \varphi$ are the Boyer-Lindquist coordinates, $\eta = Q/E^2$ and $\xi = L/E$ are 
the Chandrasekhar constants, $L$ is the angular momentum of the photon, $E$ is its energy at infinity, $Q$ is the 
Carter separation constant \cite{Carter_1968}. The auxiliary variables $R_1$ and $\theta_1$ have no physical meaning and 
are needed only for the convenience of numerical integration of the equations of motion. For the same purposes, it 
is convenient to use the radial coordinate~$R$ instead of the traditional~$r$, since the coordinate $R$ varies in a limited 
interval from zero to unity if $r$ is measured in units of the throat radius $q$.

\section{General properties of the accretion disk image}

     Earlier, in the papers \cite{Bugaev_2021, Repin_2022}, the images of a standard screen observed through the mouth of 
a wormhole and the shadow of a wormhole against the background of a standard Lambertian screen were already 
constructed. These papers show that the brightness of the image is highly non-uniform, and the image itself contains 
ring-shaped structures.  There are also graphs of the dependence of the brightness of the observed image on the impact 
parameter. In this paper, we show that the similar situation also takes place for the accretion disk observed through the throat 
of the wormhole. In this case, we will still consider the radiation source to be Lambertian, i.e., we will consider that the surface 
of the accretion disk emits uniformly in all directions.

\begin{figure*}[!htb]
  \centerline{
  \includegraphics[width=0.435\textwidth]{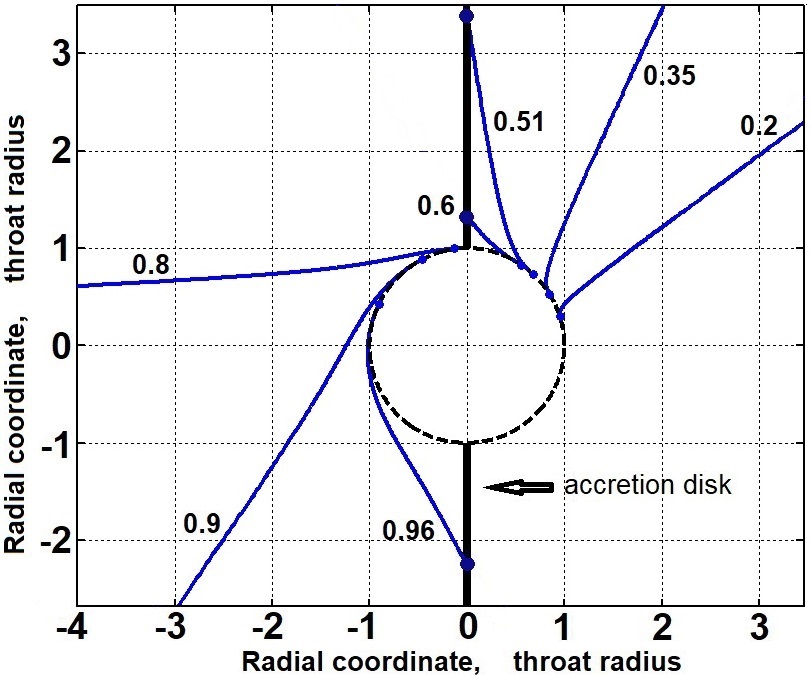}
  \includegraphics[width=0.455\textwidth]{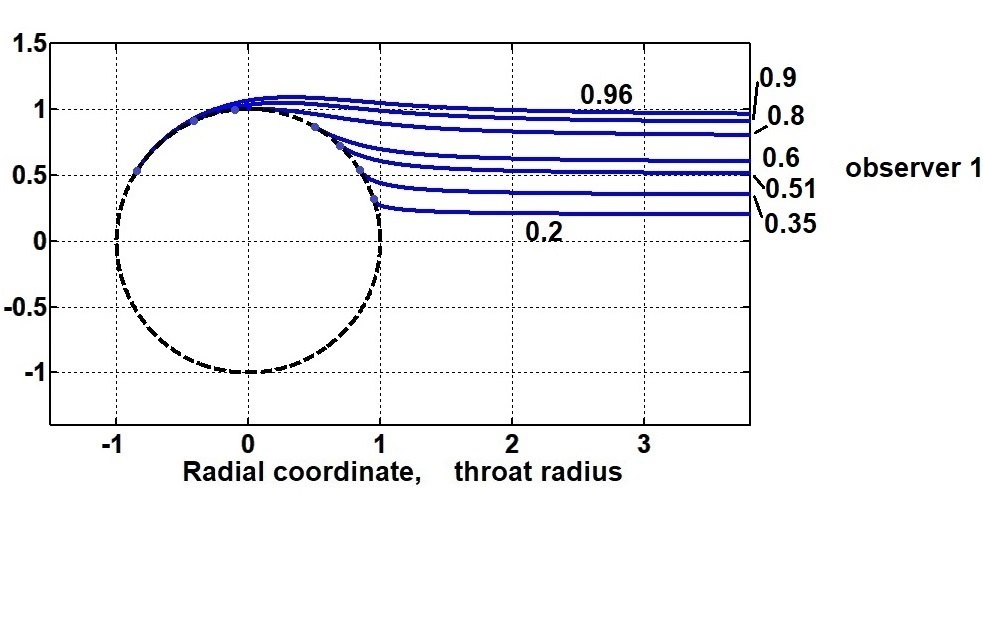}
             }
  \caption{Photon trajectories near the wormhole throat. The dotted circle marks the wormhole throat. 
                  The vertical line marks the accretion disk in space-2. The left panel shows trajectories in space-2, 
                  and the right panel shows trajectories in space-1. Each trajectory has its impact parameter next to it.}
  \label{Trajectories_7}
\end{figure*}

      Recall that due to the reversibility of the beam we can consider the motion of photons from the observer to the object.
Fig.~\ref{Trajectories_7} shows a two-dimensional cross-section of the wormhole and several trajectories of photons moving 
from the observer~1 (Fig.~\ref{Observer_and_disk}) towards the wormhole, which is indicated by the dotted circle. 
The impact parameters of all photons are indicated in the figure and their values are measured in units of the throat radius $q$.
At first the trajectories are almost parallel because the observer is far from the throat.
Part of the trajectory from the observer to the wormhole throat lies in space-1 and is shown in the right panel, and after passing 
the throat the trajectory lies in space-2 (left panel), i.e. in the same space as the accretion disk. The accretion disk itself is 
indicated by a vertical line. It follows from Fig.~\ref{Trajectories_7} that photons with impact parameters $b_1 = 0.2q$ and 
$b_2 = 0.35q$ after passing the throat do not enter the accretion disk and go to infinity. Photons with impact parameters 
$b_3 = 0.51q$ and $b_4 = 0.6q$ enter the accretion disk in space-2, but in different places. The photon with impact parameter 
$0.6q$ enters the disk closer to the throat. Starting with impact parameter $b = 0.793q$, the photon trajectory enters space-2 
after the plane of the accretion disk and goes to infinity, not entering the disk. In Fig.~\ref{Trajectories_7}, these are the 
trajectories with the impact parameters $b_5 = 0.8q$ and $b_6 = 0.9q$. Finally, with the impact parameter value of $0.940q$, 
the photon turns a total of~$270^\circ$ and then moves along the accretion disk. With a further increase of the impact 
parameter, the photon will again hit the accretion disk, but from its other side. For example, such a photon can move with an 
impact parameter $b_7 = 0.96q$ in Fig.~\ref{Trajectories_7}. If we increase the impact parameter even more, then due to 
the large rotation angle near the throat, the photon will no longer be able to move away to infinity, but will always hit the 
accretion disk from one side or the other. Due to the low brightness of this part of the image, it will be impossible to distinguish 
such \glqq rings\grqq{} from each other.

        Thus, the image of the accretion disk should consist of two concentric rings of different widths and variable intensity.

       As the disk inclination angle to the observer's line of sight changes, this picture will, of course, be distorted. This effect will 
be discussed in the next Section.

\section{Images of the accretion disk}
  
      Recall that in our model, the accretion disk is assumed to extend from the throat of the wormhole to infinity. To construct 
a single image of the accretion disk with acceptable accuracy, it is necessary to take into account the trajectories of one to 
about eight million photons, depending on the degree of detail of the image. It is also possible to separately construct the most
interesting details of the image of the accretion disk. For this purpose, it is necessary to use a denser grid for constructing 
the trajectories, keeping the number of photons approximately the same, i.e. equal to several million.

 \begin{figure*}[!hbt]
  \centerline{
  \includegraphics[width=10cm]{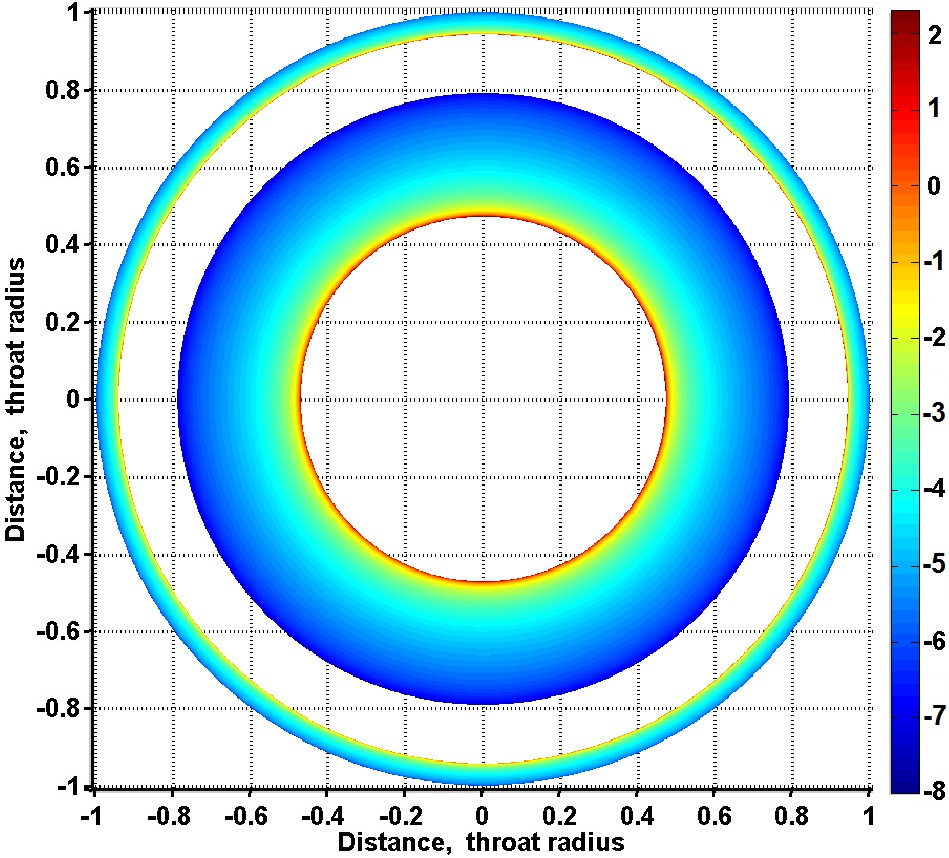}
             }
  \caption{The accretion disk as seen at the observer's polar angle $\theta = 0^\circ$. 
                The image brightness is shown in false colors and on a logarithmic intensity scale. 
                The brightness is presented in relative units.}
  \label{Disk_01}
\end{figure*}
 
      Fig.~\ref{Disk_01} shows the image of the accretion disk as seen through the wormhole throat. The observer's line of sight 
is tilted at $\theta = 0^\circ$, i.e. this image approximatly corresponds to the position of observer~1 in
Fig.~\ref{Observer_and_disk}, who observes the disk \glqq perpendicular\grqq{} to its plane. The general appearance of this 
image was discussed in the previous section. However, the objects of space-1 that this observer could see outside the wormhole 
throat are not shown in the Figure. The brightness of the image is presented in false colors and on a logarithmic intensity scale in
relative units. As follows from Fig.~\ref{Disk_01}, the accretion disk is seen as an object with highly non-uniform brightness, 
consisting of two rings, wide and narrow, of variable intensity, in which characteristic features can be distinguished. If we compare 
the brightness of different parts of the image, it follows from Fig.~\ref{Disk_01} that the brightness changes by many orders 
of magnitude.

      In the innermost part of the image there is a very thin and bright ring, which is formed by photons coming from very distant 
parts of the accretion disk. Thus, the photon coming to the observer with the impact parameter $b_1=0.48q$ was emitted by a 
point of the disk with the radial coordinate $r_1 = 12.5q$, and for the impact parameters $b_2=0.4725q$ and $b_3=0.4705q$ 
this corresponds to the points of the disk with the radial coordinate $r_2 \approx 50q$ and $r_3 \approx 240q$. This means that 
very large areas of the accretion disk are observed in a very narrow range of the impact parameter. That is why the brightest 
part of the image is located here. Note also that these photons are emitted by those parts of the accretion disk where the 
influence of relativistic effects can already be neglected. However, even in this case, the relativistic effects cannot be neglected 
when constructing the photon trajectories in the wormhole field if we construct an image of the accretion disk through the throat.

       Another bright ring can be seen at the impact distance of $b_4 = 0.940q$ from the center of the wormhole throat. The reason 
for its appearance is the same as for the first ring: here the observer sees the photons coming from the \glqq other side\grqq{} of 
the accretion disk, from very distant parts of it, and, therefore, each pixel of the image collects the photons emitted from a very 
large area.

       It is clear that when observing such an image with real astronomical instruments we will see only these bright and narrow rings, 
and other parts of the image will be suppressed. At the same time, we emphasize that the observer sees the photons coming from 
all parts of the disk, but the image itself turns out to be very non-uniform in brightness.
 
\begin{figure}[!htb]
  \centerline{
  \includegraphics[width=0.95\columnwidth]{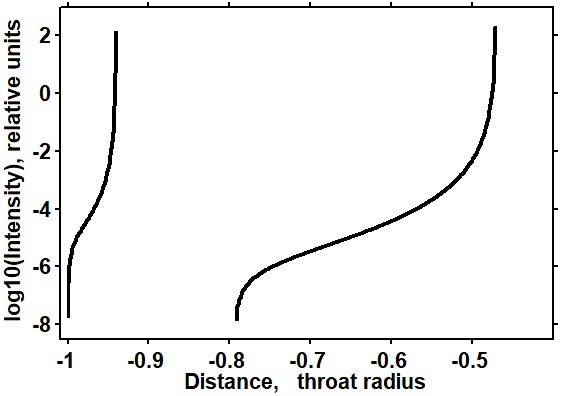}
             }
  \caption{Distribution of radiation intensity in the image of the accretion disk along the radial coordinate for the observer
                 located at the point with polar angle $\theta = 0^\circ$.}
  \label{Intensity_plot_1}
\end{figure}

\begin{figure*}[!htb]
  \centerline{
  \includegraphics[width=0.82\textwidth]{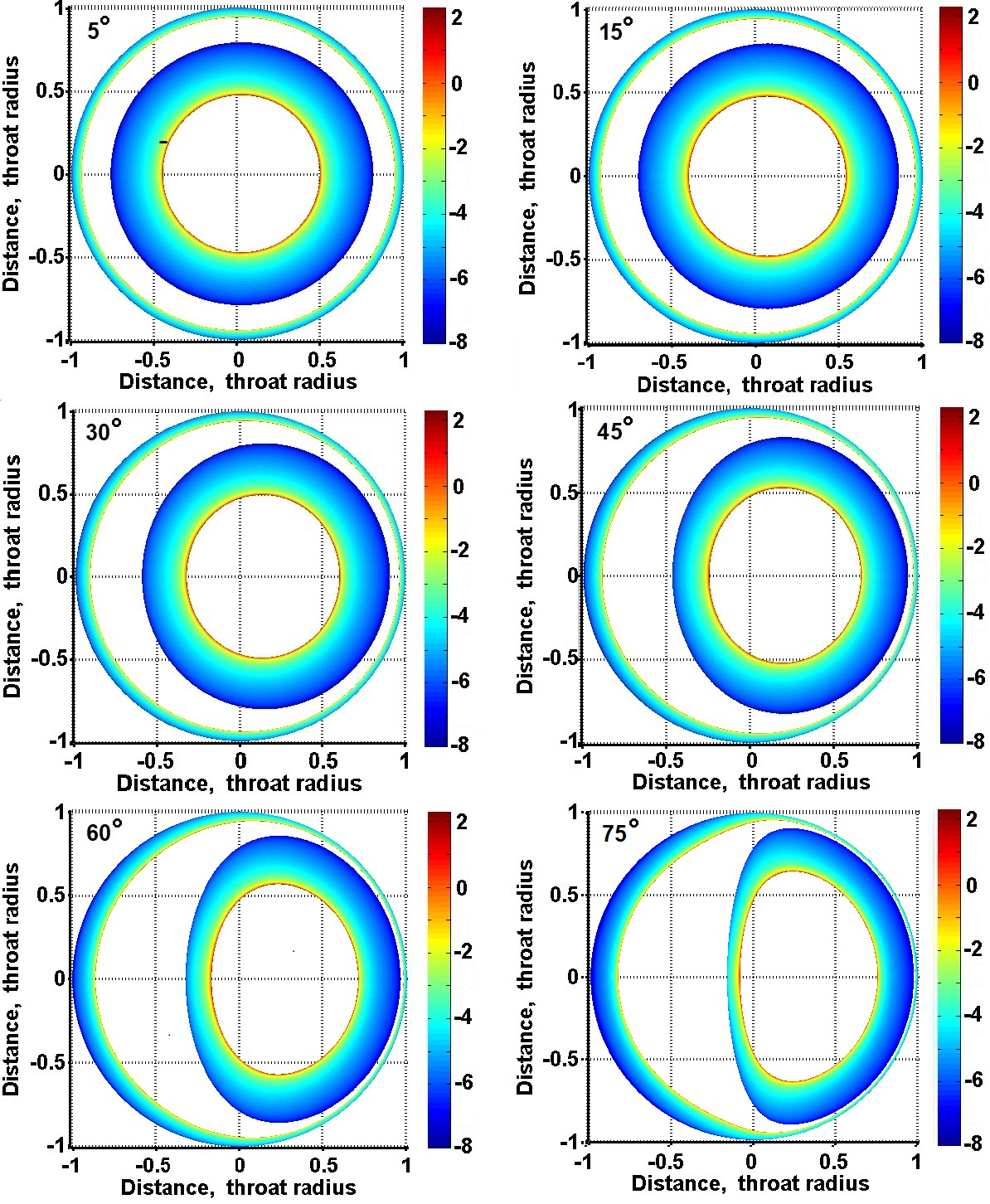}
             }
  \caption{View of the accretion disk at the angles of inclination to the observer's line of sight from ~$0^\circ$ to $75^\circ$.}
  \label{Disk_5_75}
\end{figure*}

       Fig.~\ref{Intensity_plot_1} shows a graph of the radiation intensity distribution in the equatorial section of the accretion disk 
image shown in Fig.~\ref{Disk_01}. The intensity distribution is shown only for the left part of the image, since the image is 
symmetrical. The graph clearly shows that near the image boundary, at the impact parameter value $b \approx 0.48q$, the 
intensity really increases very sharply and the graph becomes almost vertical. Theoretically, it should tend to infinity. Also, near 
the impact parameter value $b \approx 0.793q$, a very sharp drop in intensity is observed, which theoretically should tend to zero. 
It can also be noted that near the values $b = 0.64q$ and $b = 0.98q$, there are the inflection points on the graph.

       Fig.~\ref{Disk_5_75} shows other images of the accretion disk at different angles of its inclination to the observer's line of 
sight. The angle values are indicated in the upper left corner of each panel.

       As follows from Fig.~\ref{Disk_5_75}, the image becomes asymmetric when the angle between the observer's line of sight and 
the plane of the accretion disk changes. This asymmetry increases with increasing angle of inclination of the disk to the observer's 
line of sight. However, for any position of the observer relative to the disk, the beam with the impact parameter $b = 0$ does not 
hit the accretion disk when moving in space-2. This is easy to understand from general physics considerations, looking at Fig.~\ref{Observer_and_disk}. In turn, this means that in the very center of the image of the wormhole throat there cannot be 
an image of the accretion disk points. The reason for this is that in the model under consideration the disk is considered to be 
infinitely thin. With its finite thickness, the disk points will, of course, be observed in the center of the throat image.

       At large tilt angles, the image asymmetry causes the outer bright ring in one part to become so thin that it is difficult to depict 
at the correct scale. Nevertheless, the general appearance of the image is clear.

\begin{figure*}[!htb]
  \centerline{
  \includegraphics[width=0.83\textwidth]{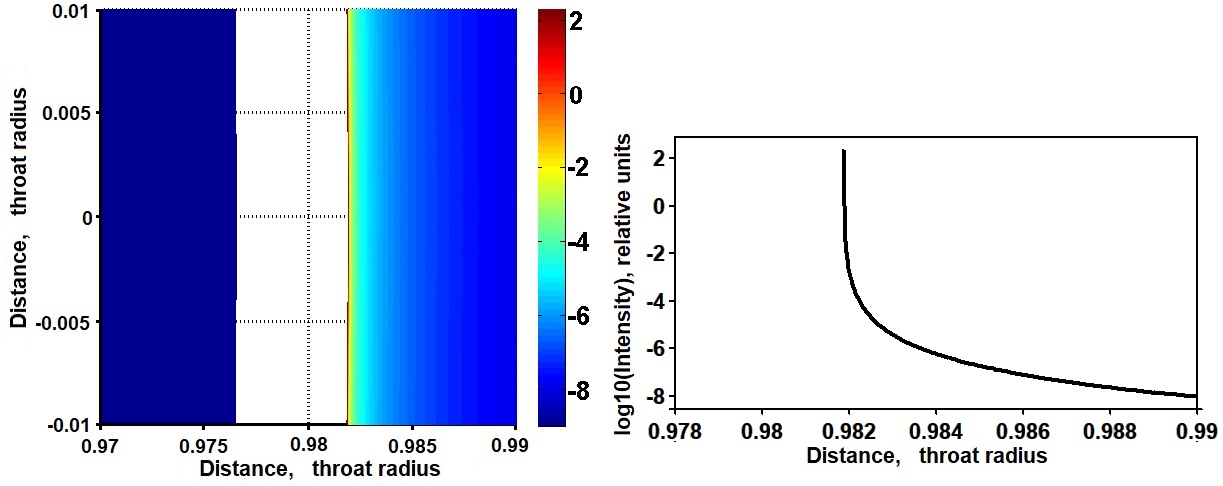}
             }
  \caption{Details of the brightness distribution of the accretion disk image as seen through the wormhole throat. 
                The disk tilt angle is $75^\circ$. 
                The brightness distribution graph is shown for the horizontal center of the image.}
  \label{Intensity_plot}
\end{figure*}

       Fig.~\ref{Intensity_plot} shows the details of the intensity distribution in a small fragment of the accretion disk image from 
$b = 0.97q$ to $b = 0.99q$ by the impact parameter at the inclination angle of $\theta = 75^\circ$. This fragment contains the 
areas of both very high and very low brightness. This fragment is located on the far right of the image in Fig.~\ref{Disk_5_75}. 
There is also an area where the accretion disk is not visible at all. The Figure also shows a plot of the intensity distribution along 
the radius in the middle of the image.

\begin{figure*}[!htb]
  \centerline{
  \includegraphics[width=0.82\textwidth]{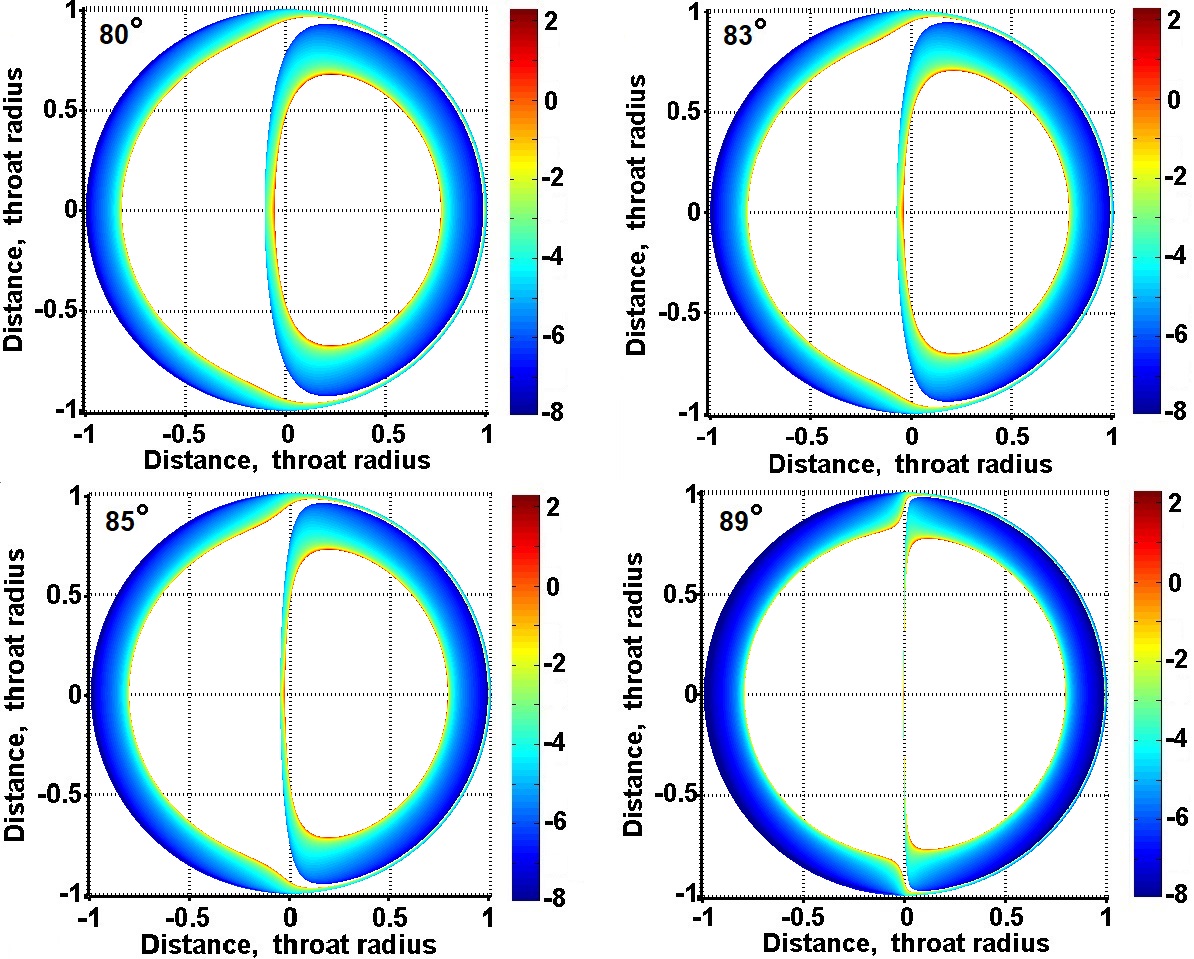}
             }
  \caption{The accretion disk as seen at angles to the observer's line of sight from $80^\circ$ to $89^\circ$. 
                The image brightness is shown in false colors and on a logarithmic intensity scale.}
  \label{Disk_80_89}
\end{figure*}

      As the disk tilt angle increases further, qualitatively new details appear in the image. Fig.~\ref{Disk_80_89} shows images 
of the accretion disk for tilt angles $\theta$ from $80^\circ$ to $89^\circ$. At so large tilt angles, i.e. when the observer looks 
almost \glqq along\grqq{} the disk surface, new details appear in the image.  In the upper and lower parts of the image, strongly 
curved bright features appear, which correspond to photons coming from very distant parts of the disk. At the same time, 
the right and left parts of the image look very similar, and the bright arc in both parts of the image approaches the already known 
value of the impact parameter $b = 0.793q$. But from general physical considerations it is clear that this should be so if 
the observer looks almost along the plane of the accretion disk. Moreover, the image of the accretion disk, constructed for the 
angle $\theta = 91^\circ$, shows that it looks exactly the same as for the angle $\theta = 89^\circ$, but it is symmetrical 
relative to the vertical axis.

         All these subtle effects are related to the degenerate model of the accretion disk, which is assumed to be infinitely thin. 
Of course, when taking into account the finite thickness of the disk, all these details are washed out.

        Such details, if they could be obtained in observations, would give a strong argument in favor of the fact that we are 
seeing an image of a wormhole, and not a black hole.

\section{Discussion}

       The images of the accretion disk shown in Fig.~\ref{Disk_01}, \ref{Disk_5_75}, \ref{Disk_80_89} are fundamentally different 
from the images of the accretion disk around the black holes. The main difference is that in this case we can observe the image 
inside the shadow of the astrophysical object. It is fundamentally impossible to observe any details inside the shadow of a black 
hole, but it is possible inside the silhouette of a traversable wormhole. A series of such images, which could be obtained by 
observing with interferometers, are presented in this paper.

       The obtained results show that when observing the accretion disk through the throat of the wormhole, there are very strong
distortions of the image of this disk. The disk itself, which is assummed to extend from the throat of the wormhole to infinity, looks 
like two rings or ovals of highly non-uniform intensity, which in different parts of the image differs by many orders of magnitude. 
Such strong distortions and differences in brightness can lead to the fact that part of the image will not be detected by 
astronomical instruments. However, to search for and identify such objects, such observations need to be carried out. These
observations need to be carried out with high angular resolution using the long-baseline interferometers \cite{Mikheeva_2020,
2022aems.conf..292M}. Similar observations have already been carried out by the Event Horizon Telescope. The selection 
of the candidates for observations can be carried out, for example, using the catalog of supermassive black holes 
\cite{Mikheeva_2019, Malinovsky_2022}.

        Constructing and studying the structure of the image of the accretion disk observed through the throat of the wormhole 
will be very important for analyzing the results of observations in future space missions \cite{Novikov_2021}. 

        The discovery of wormholes in astrophysical observations will certainly be an outstanding event in astrophysics.

\begin{acknowledgements}
     S.R. expresses his gratitude to O.N.~Sumenkova, R.E.~Beresneva and O.A.~Kosareva for the opportunity to fruitfully work 
on this problem. All authors express their gratitude to I.D. Novikov jr. for helpful discussions and assistance in preparing 
the article.

\end{acknowledgements}

% *** Astronomy and Astrophysics
\newcommand{\aaa}{Astron. and Astrophys. }
\newcommand{\aap}{Astron. and Astrophys. }
% *** Astronomical Journal
\newcommand{\aj}{Astronomical Journal }

\bibliography{Disk_in_WH_shadow}

\end{document}